\documentclass[amsmath,
twocolumn,
aps,prb]{revtex4}
\usepackage{graphicx}
\usepackage{dcolumn}

\begin{document}

\title{Cavity Enhanced Faraday Rotation of Semiconductor Quantum Dots}
\author{D. W. Steuerman}
\author{Y. Q. Li}
\author{J. Berezovsky}
\author{D. S. Seferos}
\author{G. C. Bazan}
\author{D. D. Awschalom}
\affiliation{Center for Spintronics and Quantum Computation, University of California, Santa Barbara, CA 93106}

\date{\today}
\begin{abstract}
A dielectric vertical cavity is used to study the spin dynamics of molecularly self-assembled colloidal CdSe quantum dots (QDs).  Using this structure, a nearly 30-fold enhancement of Faraday rotation is observed, which scales with the quality factor of the cavity.  In this classical nonperturbative regime, the amplified rotation is attributed to optically excited spins interacting with multiple passes of the probe photons in the cavity.  By applying this general planar cavity motif to Faraday rotation, dynamical measurements are accessible at extremely low powers on relatively small numbers of quantum confined spins.  In CdSe QDs, low power measurements reveal that contributions from exciton and electron spin precession are largely dependent upon the power of excitation.  We demonstrate that this scheme is amenable to both soft and hard systems as a means to increase detection sensitivity.
\end{abstract}
\maketitle

Advances are being made in an effort to fulfill the promise of solid state quantum computation\cite{Ekert:1996}, yet there remains a need for advanced single spin manipulation and measurement techniques.  While there has been substantial progress with both electrical\cite{Hanson:2005,Petta:2005} and optical\cite{Besombes:2004} methods, it is important to develop schemes for implementing and integrating these protocols at room temperature.  Photonic cavities provide a flexible platform to increase the sensitivity of optical measurements, as well as the possibility to study emergent light-matter interactions\cite{Purcell:1946,Ghosh:2006}.  The application of a simple multiple pass sample cell to increase sensitivity in optical techniques has existed for decades \cite{Altmann:1981}.  Recently, dielectric Bragg reflector (DBR) cavities have been exploited to investigate Faraday rotation in magnetic layers \cite{Takeda:2000}, imaging of domains in dilute magnetic semiconductors \cite{Gourdon:2003}, and time-resolved measurements in quantum wells \cite{Salis:2005}.  Many of these structures are based on an all molecular beam epitaxy fabrication approach\cite{Kavokin:1997}.
   
In this letter, we describe a generic platform that enhances the Faraday rotation (FR) of molecularly self-assembled colloidal CdSe quantum dots (QDs).  With these soft materials embedded in a dielectric vertical cavity, we use time resolved Faraday rotation (TRFR) to investigate spin dynamics and coherence of the QD films.  By independently controlling the pump and probe wavelengths, we systematically assess the role of the cavity in the resulting enhanced signal.  More than an order of magnitude increase in the FR angle is observed in the cavity as compared to control samples.  We find that the g-factors and spin coherence times of the QDs remain unaffected by introduction into the cavity.  The increase in detection sensitivity allows for the systematic investigation of spin dynamics to pump power densities of 0.5 mW/cm$^2$ where we find that the contributions to spin precession are largely power dependent.  This scheme could be incorporated into a host of systems (fluctuation QDs, planar devices, etc.) as a generic approach to increase the signal to noise ratio in FR measurements without perturbing the intrinsic properties of the system under investigation.

\begin{figure}[b]\includegraphics{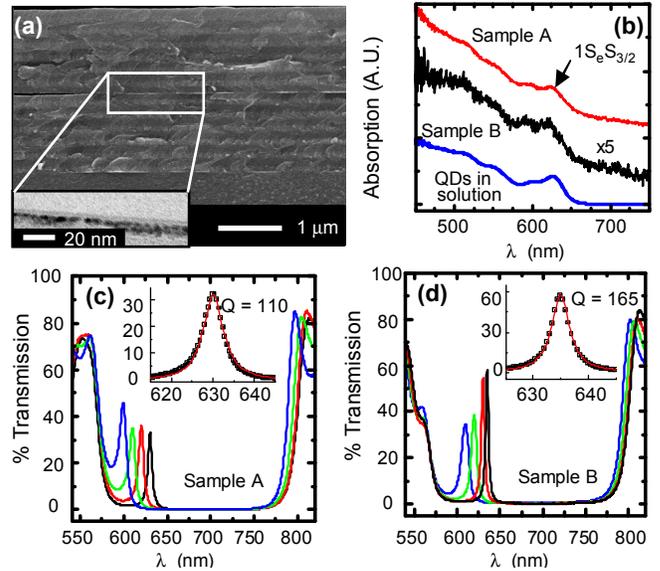}\caption{\label{fig1} 
    (a) Scanning electron micrograph of cavity cross-section.  Inset: transmission electron micrograph of embedded moleularly self-assembled CdSe QDs. (b) Absorption spectra of the control areas of both samples and QDs in toluene (spectra offset for clarity). (c) Transmission spectra of four different positions on cavity A. Inset: spectrum of cavity mode at 630 nm (squares) and the fit (line) with Q=110. (d) Transmission spectra of four different positions on cavity B. Inset: spectrum of cavity mode at 635 nm with Q=161.}\end{figure}

Fabrication of the DBR cavity begins with the electron beam evaporation of 5.5 pairs of TiO$_2$/SiO$_2$ quarter-wavelength layers deposited on a 1 mm thick glass substrate.  Next, the sample is capped with a thin SiO$_2$ layer to position a suitable surface for the molecular-assembly of QDs in the center of the completed cavity.  This SiO$_2$ surface is functionalized by immersing the sample into a 1 mM solution of 1,4-dihexyloxy-2,5-bis[4'-thiolstyryl]benzene\cite{Seferos:2004} in toluene for 2-4 hours.  This molecule contains two thiol functionalities for chemically binding CdSe QDs, and similar compounds have previously been used to fabricate and test metal-molecule-metal junctions\cite{Kushmerick:2004,Seferos:2005}.  The CdSe QD (6.8 nm diameter) layer is assembled by soaking the modified substrate in a QD solution for 13-16 hours\cite{Evident}. The same molecular self-assembly procedures are repeated a few times until a sufficiently high optical density (OD = 0.01) is obtained.  This assembly protocol allows the OD of each sample to be varied by adjusting the number of deposition cycles.  For this study we have created one sample with low OD (Sample A) and one with high OD (Sample B), $\textit{vide infra}$.  QD modified samples were subsequently coated with a wedge-shaped SiO$_2$ layer to tune the half-wavelength layer, and therefore the cavity mode resonance continuously.  Finally, a second TiO$_2$/SiO$_2$ DBR that is identical to the first series of depositions is evaporated to complete the cavity structure as can be seen by the electron microscopy cross-sections in Fig. 1(a).  The inset shows that the film consists of 1-3 layers of QDs.  

In order to assess the role of the cavity, it is critical to compare the response of the identical film inside and outside of the structure.  This is achieved by dividing each substrate into a cavity region and a control region.   The control side is masked during depositions of the bottom and top DBRs, whereas the whole substrate is exposed to the QD assembly protocol.  In order to ensure that the films are identical, a thin layer of SiO$_2$ is deposited in the control region before and after QD deposition.  This procedure allows for quantitative comparison between the film in the cavity versus the control.  Shown in Fig. 1(b) are the optical absorption spectra of the control region of both samples (A and B).  The first exciton absorption (1S$_e$-1S$_{3/2}$ transistion) appears at 625 nm in both films and a reference solution of QDs in toluene.  The linker molecules used in the assembly process do not have absorption for $\lambda$ $\geq$ 450 nm\cite{Seferos:2004}.  

In the cavity structures, a single DBR typically has a maximum reflectivity of 96$\%$ at the center wavelength, 685 nm, and stop-band edges of 550 nm and 800 nm.  The refractive indices for SiO$_2$ and TiO$_2$ layers are estimated to be 1.46 and 2.07, respectively.  Using existing analysis\cite{Babic:1992}, we find the phase penetration depth to be 270 nm.  Optical characteristics of a complete cavity are shown in Fig. 1(c).   The transmission spectra of four different positions of sample A demonstrate that the cavity mode can be tuned from 595 nm to 630 nm.  The inset shows the width of the resonance and its fit to a lorentzian line-shape.  A similar analysis is conducted for sample B in Fig. 1(d).  Here the cavity mode can be tuned from 600 nm to 635 nm and has a higher Q due to reduced absorption from the embedded QDs.  When these cavities are compared to similar structures without the QD layer, the maximum transmission at a cavity mode drops from 80$\%$ to 54$\%$.  We also find that the quality factors of these structures are similar implying that a high quality interface persists on both sides of the QD layer despite the vast change in surface chemistry.  

\begin{figure}[b]\includegraphics{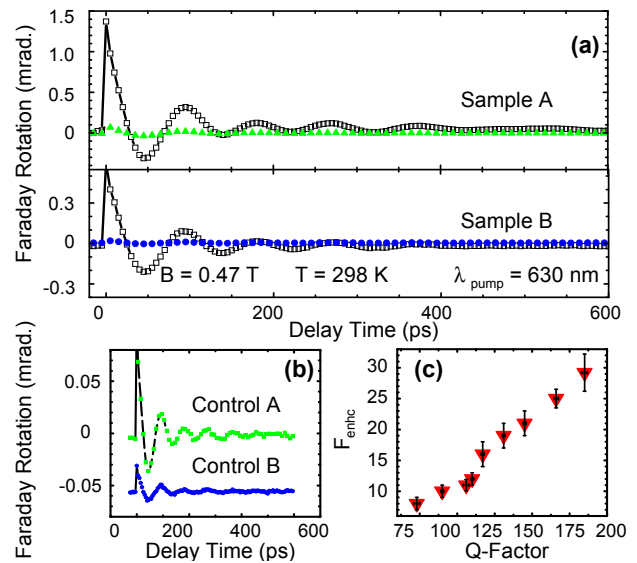}\caption{\label{fig2} 
    (a) Typical TRFR scans of both cavities in resonance with probe(open symbols) and the corresponding controls measured with identical conditions (closed symbols).  (b) Zoom-in of both control measurements.  (c) Cumulative measurements of TRFR as function of position over both samples results in F$_{enhc}$ as a function of Q-factor.}\end{figure}
        
We measure coherent spin dynamics by time-resolved Faraday rotation in the Voigt geometry \cite{Crooker:1997}.  A regeneratively amplified Ti:sapphire laser pumps two optical parametric amplifiers at a repetition rate of 250 kHz to produce a pair of synchronized  independently tunable optical pulses of duration 200 fs (pump and probe with wavelength  $\lambda_{pump}$ and $\lambda_{probe}$, respectively).  The helicity of the circularly polarized pump beam is varied by a photoelastic modulator at 42 kHz, whereas the linearly polarized probe beam is mechanically chopped at a frequency of 390 Hz.  At room temperarture, these pulses are focused to a 100 $\mu$m spot on the sample which resides between two permanent magnets used to apply a field of 0.47 T perpendicular to the pump and probe direction.  The magnetization and subsequent precession of the excited carriers results in the rotation of the linear polarization of the probe which is detected with a balanced photodiode bridge via two cascaded lock-in amplifiers and a voltage pre-amplifier.  The relative time delay between the pump-probe pulses is adjusted by a mechanical delay line.  Probe powers are typically 1 mW/cm$^2$ and all results were independent of probe intensity.  

A typical TRFR scan of sample A with a cavity resonance of 630 nm, P$_{pump}$ = 40 mW/cm$^2$ and $\lambda_{pump}$= 560 nm is plotted in Fig. 2(a).  The signal from the cavity is signifigantly larger than that of the control measured under the same conditions.  Since $\lambda_{pump}$ is below the stop band edge, the low reflectivity of the DBRs makes any effect from the cavity on the pump beam negligible.  Hence, the enhancement observed here can be solely attributed to the cavity's effect on the resonant probe beam.  In order to characterize our samples more quantitatively we fit the TRFR data to the following equation:

\begin{eqnarray}
\theta_F(t)
&=& \theta_1exp(-t/T_{21}^*)cos(2\pi\nu_1t+\phi_1) \nonumber\\
&+& \theta_2exp(-t/T_{22}^*)cos(2\pi\nu_2t+\phi_2) \nonumber\\
&+& \theta_0exp(-t/t_0)+\theta_C
\end{eqnarray}

where t is the delay time of the pump pulse, $\theta_1$ ($\theta_2$) the 
amplitude of the first (second) spin precession component, T$_{21}$$^*$ (T$_{22}$$^*$) the transverse spin coherence time, $\nu_1$ ($\nu_2$) the Larmor frequency, $\phi_1$ ($\phi_2$) the phase, $\theta_0$ and t$_0$ the amplitude and decay time constant respectively for the non-oscillating background, and $\theta_C$ is the offset.  Analysis of sample A yields $\theta_1$=0.12 mrad and $\theta_2$=0.55 mrad for the cavity at a resonance of 630 nm.  In contrast, the two FR amplitudes for the control are found to be $\theta_1$=0.011 mrad and  $\theta_2$=0.046 mrad.  The Larmor frequencies are found to be $\nu_1$=7.8 GHz,  $\nu_2$=10.8 GHz, corresponding to g-factors g$_1$=1.18 and g$_2$=1.64, respectively.  The dephasing times T$_{21}$$^*$ and T$_{22}$$^*$ are found to be approximately 100 ps and 250 ps, respectively.  We find no discernible difference between g-factors or dephasing times of the cavity and the control.  The origin of the two distinct components of FR oscillations has previously been attributed to electron and exciton precession and that picture will be adopted here\cite{Gupta:2002,Stern:2005}.  The important difference between the control and the cavity resides in the increased FR signal.  A similar analysis for sample B results in a larger enhancement.

\begin{figure}[b]\includegraphics{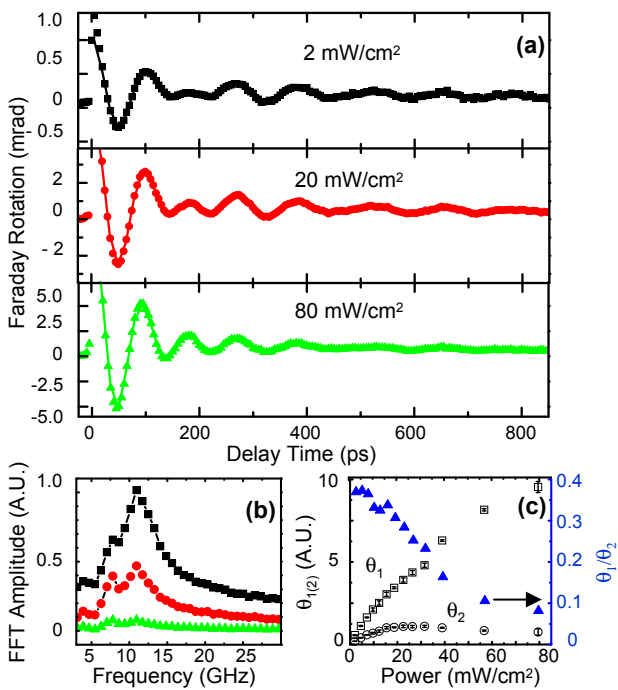}\caption{\label{fig3} 
    (a) Three representative TRFR scans at various powers.  (b) FFT of the three TRFR scans.  (c) Amplitudes of precession frequencies as a function of power using Eq. 1.  }\end{figure}

In order to quantify the increase in Faraday rotation we define an enhancement factor using coefficients from Eq (1).  Since the amplitude associated with the electron Larmor frequency is small for the control region, we use the ratio between the sums of the two FR amplitudes to calculate the cavity enhancement factor,  i.e. F$_{enhc}$=($\theta_1$+$\theta_2$)$_{cavity}$/($\theta_1$+$\theta_2$)$_{control}$.  Averaging of the scans over at least five positions in the control and two in the cavity yields F$_{enhc}$=12$\pm$1 for the sample A and F$_{enhc}$=20$\pm$2 for sample B at a 630 nm resonance.  To further explore this, we made measurements as a function of position on both cavities.  This allows for the dependence of F$_{enhc}$ on the resulting quality factor to be determined.  Measurements with probe wavelengths between 620-625 nm are avoided because the signal from the control is quite weak and FR oscillations vanish near the peak of the 1S$_e$-1S$_{3/2}$ transition.  As can be seen in Fig. 2(c) the enhancement is dependent on the quality factor, and in our case an increase of nearly 30 times is possible with a Q of 185.        

This phenomenon can be explained by considering how the probe beam interacts with the cavity, and subsequently the QDs.  This type of cavity structure can be thought of as an environment that supports multiple passes of the probe beam through the QD film or alternatively, an optical localization picture where interference effects lead to cancellation and amplification of the optical field.  This results in an antinode at the center of the cavity.  Either approach leads to an enhanced interaction between pump excited QDs and the probe beam, thus allowing for improved detection of Faraday rotation.  In order to determine the number of roundtrips the probe beam travels in the cavity, one can compare the transmission on resonance with a QD loaded cavity to one that is unloaded.  In cavity B at a resonance of 630 nm, these values are 54$\%$ and 80$\%$ respectively, while the OD of the film is 0.006.  This implies that the probe photons pass through the QD film 29 times.  As expected of any linear optical process, we find these conditions enhance the FR by a factor of 28.5.  This is borne out in Fig 2(c) when one considers that Q is linearly proportional to the photon lifetime or number of roundtrips in the cavity.                              

The cavity enhancement can be used to study QDs at extremely low pump and probe power levels.  As shown in Figs. 3(a), the spin dynamics at low pump power (2.5 mW/cm$^2$) are quite different from higher power in that, the beating of FR oscillations is much stronger at low power.  In Fig. 3(b), Fast Fourier transforms (FFTs) of time-domain TRFR data reveal that the relative weight of the low frequency component is greater at lower power, and the g-factors are independent of pump power.  Once again, fitting the FR data with Eqn.(1) demonstrates that the two precession components of FR vary as a function of pump power (P$_{pump}$).  As can be seen in Fig. 3(c), both $\theta_1$ and $\theta_2$ increase with P$_{pump}$, but the low frequency component $\theta_1$ begins to saturate much faster than  $\theta_2$.  Qualitatively similar results were observed for various $\lambda_{pump}$ and $\lambda_{probe}$ conditions.  

Previous studies have probed large numbers of QDS either in polymer glasses\cite{Gupta:2002} or thick densely packed films\cite{Stern:2005}.  Using both electron microscopy and optical absorbance, we find that our films have a density of 1.3 x 10$^{12}$ QDs/cm$^{2}$ and correspondingly we probe 9 x 10$^8$ QDs.  If we assume signal levels observed in our control films as a lower bound and reduce our spot size with a high numerical objective, we conservatively estimate the current detection limit of Faraday rotation to 3 x 10$^4$ QDs using this cavity.      

In conclusion, we have demonstrated a flexible platform to increase the sensitivity of ultra-fast spin dynamics measurements.  We have established that this design is compatible with soft materials systems and may prove useful as molecules begin to play an increasingly important role in spintronics.  Finally, we take advantage of the cavity enhanced signal to show that contributions to FR are pump power dependent in CdSe QDs.  Single spin detection may be within experimental reach with the application of this cavity structure to other quantum confined spin systems.

We thank Dr. Jan Lofvander for his assistance in the fabrication of electron microscopy samples, and acknowledge
support from DARPA, ONR, and NSF.


\begin{thebibliography}{0}
\expandafter\ifx\csname natexlab\endcsname\relax\def\natexlab#1{#1}\fi
\expandafter\ifx\csname bibnamefont\endcsname\relax
  \def\bibnamefont#1{#1}\fi
\expandafter\ifx\csname bibfnamefont\endcsname\relax
  \def\bibfnamefont#1{#1}\fi
\expandafter\ifx\csname citenamefont\endcsname\relax
  \def\citenamefont#1{#1}\fi
\expandafter\ifx\csname url\endcsname\relax
  \def\url#1{\texttt{#1}}\fi
\expandafter\ifx\csname urlprefix\endcsname\relax\def\urlprefix{URL }\fi
\providecommand{\bibinfo}[2]{#2}
\providecommand{\eprint}[2][]{\url{#2}}

\end{thebibliography}


\begin{thebibliography}{18}
\bibitem{Ekert:1996} A. Ekert, and R. Jozsa, Rev. Mod. Phys. \textbf{68}, 733 (1996).
\bibitem{Hanson:2005}R. Hanson, L. H. W. van Beveren, I. T. Vink, J. M. Elzerman, W. J. M. Naber, F. H. L. Koppens, L. P. Kouwenhoven, L. M. K. Vansersypen, Phys. Rev. Lett \textbf{94}, 196802 (2005). 
\bibitem{Petta:2005} J. R. Petta, A. C. Johnson, J. M. Taylor, E. A. Laird, A. Jacoby, M. D. Lukin, C. M. Marcus, M.P. Hanson, and A. C. Gossard,  Science \textbf{309}, 2180 (2005).
\bibitem{Besombes:2004} L. Besombes, Y. Le´ger, L. Maingault, D. Ferrand, H. Mariette, and J. Cibert, Phys. Rev. Lett.  \textbf{93}, 27403 (2004).
\bibitem{Purcell:1946} E. M. Purcell, Phys. Rev. B \textbf{69}, 681 (1946).
\bibitem{Ghosh:2006} S. Ghosh, W. H. Wang, F. M. Mendoza, R. C. Myers, X. Li, N. Samarth, A. C. Gossard, and D. D. Awschalom, /cond-mat/0509500 to appear in Nature Materials.
\bibitem{Altmann:1981} J. Altmann, R. Baumgart, and C. Weitkamp, Appl. Opt. \textbf{20}, 995 (1981).
\bibitem{Takeda:2000} E. Takeda, N. Todoroki, Y. Kitamoto, M. Abe, M. Inoue, T. Fujii, and Ken'ichi Arai, J. Appl. Phys. \textbf{87}, 6782 (2000).
\bibitem{Gourdon:2003} C. Gourdon,  V. Jeudy, M. Menant, D. Roditchev, Le Anh Tu, E. L. Ivchenko, and G. Karczewski, Appl. Phys. Lett. \textbf{82}, 230 (2003).
\bibitem{Salis:2005} G. Salis and M. Moser, Phys. Rev. B \textbf{72}, 115325 (2005).
\bibitem{Kavokin:1997} A. V. Kavokin, M. R. Vladimirova, M. A. Kaliteevski, and O. Lyngnes, Phys. Rev. B \textbf{56}, 1087 (1997).
\bibitem{Seferos:2004} D. S. Seferos, D. A. Banach, N. A. Alcantar, J. N. Israelachvili, and G. C. Bazan, J. Org. Chem. \textbf{69}, 1110 (2004)
\bibitem{Kushmerick:2004}  J. G. Kushmerick, J. Lazorcik, C. H. Patterson, R. Shashidar, D. S. Seferos, and G. C. Bazan, Nano Lett. \textbf{4}, 639 (2004) 
\bibitem{Seferos:2005} D. S. Seferos, S. A. Trammell, G. C. Bazan, and J. G. Kushmerick, Proc. Natl. Acad. Sci. U. S. A. \textbf{102}, 8821 (2005).
\bibitem{Evident} CdSe core dots are purchased from Evident Technology.
\bibitem{Babic:1992} D. I. Babic and S. W. Corzine, IEEE J. Quantum Electron. \textbf{28}, 514 (1992).
\bibitem{Crooker:1997} S. A. Crooker, J. J. Baumberg, F. Flack, N. Samarth, and D. D. Awschalom, Phys. Rev. B \textbf{56}, 7574 (1997).
\bibitem{Gupta:2002} J. A. Gupta, D. D. Awschalom, A. L. Efros, and A. V. Rodina, Phys. Rev. B \textbf{66}, 125307 (2002).
\bibitem{Stern:2005}N. P. Stern, M. Poggio, M. H. Bartl, E. L. Hu, G. D. Stucky, and D. D. Awschalom Phys. Rev, B \textbf{72}, 161303R (2005). 
\end{thebibliography}
\end{document}